\documentclass[fleqn,usenatbib,useAMS]{mnras}
\usepackage{epsfig,amssymb,amsmath,rotating,lineno,graphicx,hyperref}
\usepackage[T1]{fontenc}
\usepackage{ae,aecompl,arydshln}
\usepackage{mathtools}

\newcommand\swift{{\it Swift}}

\newcommand\xte{{\it RXTE}}

\newcommand\iue{{\it IUE}}
\newcommand\hst{{\it HST}}

\newcommand{\siv}{Si~{\sc iv}}
\newcommand{\civ}{C~{\sc iv}}
\newcommand{\heii}{He~{\sc ii}}
\newcommand{\nv}{N~{\sc v}}
\newcommand{\me}{$\dot{m}$/${\dot{m}_{Edd}}$}
\title[Reverberation in NGC~7469]{Evidence for variability timescale dependent UV/X-ray delay in Seyfert 1 AGN NGC~7469 }

\author[Pahari et al.]{Mayukh Pahari$^{1}$, I. M. M$^{\rm c}$Hardy$^{1}$, Federico Vincentelli$^{1}$, Edward Cackett$^{2}$,  
\newauthor Bradley M Peterson$^{3,4}$, Mike Goad$^{5}$, Kayhan G\"{u}ltekin$^{6}$, Keith Horne$^{7}$\\
$^{1}$ School of Physics \& Astronomy, University of Southampton, Highfield campus, Southampton SO17 1BJ, UK \\
$^{2}$ Department of Physics and Astronomy, Wayne State University, 666 W. Hancock Street, Detroit, MI 48201, USA \\
$^{3}$ Department of Astronomy, The Ohio State University, 140 W 18$^{th}$ Avenue, Columbus, OH 43210, USA \\
$^{4}$ Space Telescope Science Institute, 3700 San Martin Drive, Baltimore, MD 21218, USA \\
$^{5}$ University of Leicester, Department of Physics and Astronomy, Leicester, LE1 7RH, UK \\
$^{6}$ Department of Astronomy, University of Michigan, 500 Church Street, Ann Arbor, MI 48109, USA \\
$^{7}$ School of Physics and Astronomy, University of St. Andrews, Fife KY16 9SS, Scotland, UK}

\begin{document}

\pagerange{\pageref{firstpage}--\pageref{lastpage}} \pubyear{2019}

\maketitle

\label{firstpage}

\begin{abstract}
Using a month-long X-ray lightcurve from \xte{}/PCA and 1.5 month long UV continuum lightcurves from \iue{} spectra in 1220$-$1970~\AA, we performed a detailed time-lag study of the Seyfert 1 galaxy NGC 7469. Our cross-correlation analysis confirms previous results showing that the X-rays are delayed relative to the UV continuum at 1315~\AA ~by 3.49 $\pm$ 0.22 days which is possibly caused by either propagating fluctuation or variable comptonisation. However, if variations slower than 5 days are removed from the X-ray lightcurve, the UV variations then lag behind the X-rays variations by 0.37$\pm$0.14 days, consistent with reprocessing of the X-rays by a surrounding accretion disc. A very similar reverberation delay is observed between \swift{}/XRT X-ray and \swift{}/UVOT UVW2, U lightcurves. Continuum lightcurves extracted from the \swift{}/GRISM spectra show delays with respect to X-rays consistent with reverberation. Separating the UV continuum variations faster and slower than 5~days, the slow variations at 1825~\AA~lag those at 1315~\AA~by $0.29\pm0.06$~days, while the fast variations are coincident ($0.04\pm0.12$~day). The UV/optical continuum reverberation lag from \iue{}, \swift{} and other optical telescopes at different wavelengths are consistent with the relationship: $\tau \propto \lambda^{4/3}$, predicted for the standard accretion disc theory while the best-fit X-ray delay from \xte{} and \swift{}/XRT shows a negative X-ray offset of $\sim$0.38 days from the standard disc delay prediction.     
\end{abstract}

\begin{keywords}
accretion, accretion disc --- galaxies: Seyfert --- black hole physics --- X-rays: galaxies --- galaxies: individual: NGC~7469
\end{keywords}

\section{Introduction}

Emission-line reverberation mapping \citep{bl82,pe14}, based on measured lags between continuum and emission-line bands and the width of the emission line, is a very successful technique for determining AGN broad line region (BLR) size and black hole virial mass. Over 60 masses have currently been measured \citep{pe04,be09,be15}. Over the last 2 decades, considerable observational effort has also been put into continuum reverberation mapping, measuring the lags between a short-wavelength band, often the X-rays, and longer-wavelength UV and optical bands. The initial aim was to map the temperature structure of the accretion disc and hence find a standard candle by which distances could be estimated and the Hubble constant derived \citet{ca07}. Most such studies assumed a disc with the temperature structure as derived by \citet{sh73}. Incident high energy emission will enhance the existing thermal emission leading to a wavelength ($\lambda$) dependent lag, $\tau$, between the incident high energy, and re-radiated UV/optical emission, of $\tau \propto (M^{2}\dot{m}_{E})^{1/3} \lambda^{\beta}$ where $\beta=4/3$, $M$ is the black hole mass and $\dot{m}_{E}$ is the accretion rate in Eddington units. Initial studies \citep{co99,ca07} were consistent with $\beta=4/3$, but included only optical bands and did not extend to the X-ray bands.

Coordinated observations, usually with Rossi X-ray Timing Explorer (\xte{}) \citep{ar09,br09,ar08,ma08,ut03,mc03,sh03}, and ground-based optical telescopes, mostly revealed a good correlation, with the optical lagging behind the X-rays, consistent with the expectations of reprocessing. However the lag measurements $\sim1 \pm 0.5$ days were rarely statistically significant and could not rule out that the X-rays might lag behind the optical. These long \xte{}-based programmes, which in some cases covered up to 10 years \citep{br10}, also showed that although there was a good correlation between the X-rays and the optical bands on short timescales (weeks-months), on longer timescales (months-years) there were often trends in the optical lightcurves with no counterparts in the X-ray lightcurves.

Lags in the opposite sense, where the hard band lags the soft, are also seen on longer timescales. These hard lags have been seen in both X-ray binaries (from milliseconds to seconds) and AGNs (from days to months) \citep{pa01,mc04,ar06}, and are thought to arise due to the inward propagation on viscous timescales of mass accretion fluctuations in the disc that are then transmitted to the corona \citep{ko01,ar06,ut11}.

More intense, multi-band, observations with \swift{} \citep{sh14,mc14,ed15,fa16,tr16,ed17,mc18,pa18,ed19} confirmed the general picture of wavelength dependent UV/optical lags, consistent with disc reprocessing, although the measured lags were $\sim2-3$ times longer than expected theoretically \citep{mc14}. This discrepancy may indicate an inhomogeneous disc \citep{de11}. The \swift{} observations also provide evidence of reprocessing of high energy emission from a larger reprocessor than just the accretion disc, probably the broad line region (BLR) clouds \citep{ko19,ch19,la18,mc18,ca18,su18,pa18}. This evidence is in the form of an excess lag in the U-band \citep{ed17,fa16,ed15}, which contains the Balmer continuum, \citep{ko01} and an excess lag at 3634 \AA~ (known as the Balmer jump), and also in the fact that the reprocessing function required to explain the optical emission as reprocessing of X-ray emission, has a tail to long delays (a few days) as well as a sharp peak at short timescales ($\sim$hours) from the disc. The \swift{} observations also show that although wavelength-dependent lags following roughly $\tau \propto ~\lambda^{4/3}$ apply in most AGN between the UV and optical bands, the lag between the X-ray and UV band is usually much larger than expected purely from extrapolation of the UV-optical lag spectrum down to X-ray wavelengths \citep{da10,mo10,mo13,ed15,fa16,mc18}. Moreover, the X-ray/UV correlation is weaker than the UV/optical one \citep{ed19}. 

There are a number of possible explanations for this increased lag, including that the lag corresponds to the thermal timescale due to the thermal reverberation from a significantly hot disc with low accretion rate \citep{ka19,su18}, hot accretion flow with a disc truncation \citep{no16}, very large area of the reprocessing site \citep{pa17,pa18}, non-blackbody nature of the emerging disc spectra due to the low atmospheric density \citep{ha18}. Other potential explanations include that the X-rays do not directly illuminate the outer disc but are first reprocessed by, and scattered through, the scattering atmosphere \citep{na96}, the inflated inner edge of the accretion disc, which introduces an additional lag \citep{ga17}.

The increase in X-ray/UV-optical correlation strength on short timescales, originally noted by \citet{br09}, has also been noted in \swift{} observations. For example in NGC~5548 when variations on timescales longer than 20 days are removed from UV lightcurves, the correlation improves and, moreover, the X-ray to UV lag now falls on an extrapolation of the UV-optical lag spectrum. A similar behaviour is seen in NGC~4593 \citep{mc18} where the X-ray/UV lag decreases when long timescale variations ($>10$ days), presumably from the BLR, are removed from the UV lightcurves. The resultant X-ray/UV lag is then again in agreement with an extrapolation of the UV-optical lag spectrum to X-ray wavelengths.

The one notable exception to the general scenario that the UV/optical variations on short timescales are mainly driven by reprocessing of high energy (i.e. X-ray) variations is NGC~7469, a bright, infrared luminous, Sb-type spiral Seyfert 1 galaxy at a red-shift of 0.016268. The mass of the central supermassive black hole is $9.04^{+1.06}_{-0.97} \times 10^{6}$ \(\textup{M}_\odot\)\footnote{This neglects a further $\sim0.4$~dex uncertainty due to using the population mean value of the dimensionless factor $f$, which depends on the uncertain geometry and orientation
of the BLR in each object.} \citep{pe14a,zu11}.

\begin{figure*}
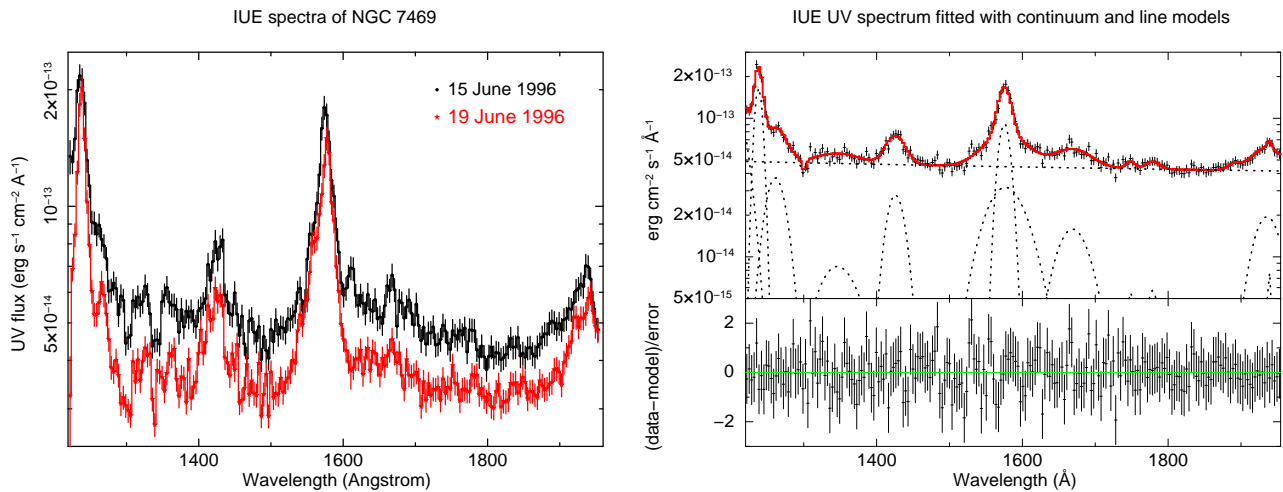

\centering
\includegraphics[scale=0.33,angle=-90]{fig1a.ps}
\includegraphics[scale=0.33,angle=-90]{fig1b.ps}
\caption{{\it UV variability of NGC 7469 and spectral modelling:} Left: \iue{} spectra of the ~NGC 7469 nucleus in the wavelength range of 1220-1970 \AA ~as observed on 15 June 1996 (circles) and 19 June 1996 (stars) respectively. A substantial decrease is observed in the UV continuum flux level within 4 days. Right: the best-fit time-averaged UV spectrum (1220-1970 \AA) from \iue{} fitted with a model (red) consisting of a continuum, broad and narrow emission features and narrow absorption (dotted lines; top) and the residual of the fitting (bottom). Each of 218 spectra is fitted separately with the best-fit model to obtain the UV continuum and line fluxes from each pointing.}
\label{iuespec}
\end{figure*}

In June-July 1996 NGC~7469 was observed almost continuously, Earth occultations excepted, for a period of $\sim$46 days by \iue{}, providing UV spectra from which lightcurves can be produced in a variety of bands \citep{wa97}. For 30 of these days there was almost continuous \xte{}/PCA monitoring. Both the \iue{} and \xte{} lightcurves are dominated by a small number of large amplitude, quasi-sinusoidal, variations with peak to peak timescales of around 15-20 days \citep[see Figure 2 of][]{na98}. 
\citet{na98} show that the peaks in the UV lightcurve led the peak in the X-rays by $\sim$4 days. Although there are only 2 cycles of variability in this analysis, these observations nonetheless led to much speculation regarding physical mechanisms which might explain the 4d X-ray lag. \citet{na98} suggested that the X-rays might be produced by up-scattering of UV photons by a variable coronal structure. Using the same UV/X-ray observations, \citet{pet04} performed UV and hard X-ray joint spectral fitting and found an anti-correlation between the UV flux and the X-ray coronal temperature. The explanation of such an anti-correlation requires strong variability in coronal structure over days rather than a simple disc-corona structure \citep{pet04}. 

However, just within the \iue{} band, \citet{wa97} estimated the delay of Ly$\alpha$, \civ{}, \nv{}, \siv{} and \heii{} emission lines with respect to the UV continuum as 2.3-3.1 days, $\sim$2.7 days, 1.9-2.4 days, 1.7-1.8 days and 0.7-1 day respectively which is broadly consistent with observations of other Seyfert galaxies. Using concurrent ground-based spectro-photometric monitoring of NGC~7469, \citet{co98} found that continuum variations at 4865~\AA~and 6962 \AA~ lag those at 1315~\AA~by 1.0$\pm$ 0.3 days and 1.5 $\pm$ 0.7 days respectively.
They also noted that the continuum variations at 1485-1825~\AA~lag those at 1315~\AA~by 0.21-0.35 days, which is consistent with the expectations from disc reprocessing. Later, using a more sophisticated spectral modelling approach, using the \hst{}/FOS spectrum as a template, \citet{kr03} found that the continuum variations at 1485~\AA, 1740~\AA~and 1825~\AA~is delayed relative to the shorter UV continuum at 1315~\AA~by 0.09, 0.29, 0.36 days respectively, again in good agreement with disc reprocessing.

The main remaining unexplained problem, therefore, is the relationship between the X-ray and UV variations. In this paper we re-examine the relationship between the X-ray and UV variations as observed by \xte~and \iue. In addition to examining the correlation in the raw lightcurves, we also search for a correlation in lightcurves from which the long timescale, large amplitude, variations have been removed (Section 3). Here we find lags that are more consistent with the reprocessing scenario (Section 4.1). We also examine archival \swift{} data which, although with considerably greater uncertainties than in the \iue ~data, allow us to extend our lag measurements into the optical bands (Section 5). These data also allow us to determine whether the apparent lag of the UV by the X-rays seen with \iue~and \xte, when considering long timescale variations, is a common phenomenon. This does not appear to be the case. We compare the observed wavelength-dependent lag with that predicted We conclude the paper (Section 7) with a brief summary of the observational results and with some general overall interpretations.

\begin{figure*}
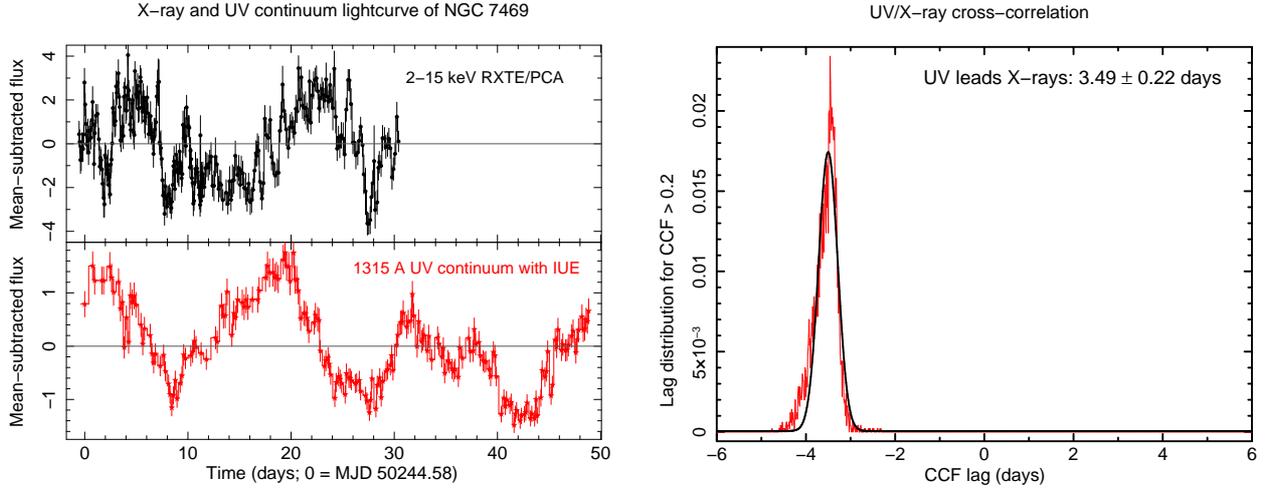

\centering
\includegraphics[scale=0.33,angle=-90]{fig2a.ps}
\includegraphics[scale=0.33,angle=-90]{fig2b.ps}
\caption{{\it X-ray/UV cross-correlation:} Left: 2-15 keV background-subtracted \xte{}/PCA lightcurve (top) and rest-frame 1315~\AA ~UV continuum lightcurve observed with \iue{}~(bottom). Both lightcurves are mean-subtracted for a better visibility of the relative flux variability. When both lightcurves are cross-correlated using Monte Carlo-based combined flux randomisation and random subset selection methods(FR/RSS), the relative frequency distribution of the cross-correlation function centroids (CCF; when cross correlation coefficient > 0.2) is shown in the right panel. The lag distribution is fitted with a Gaussian. The centroid and the FWHM/2 of the Gaussian with stronger peak is quoted as the time delay and its uncertainty, respectively. }
\label{xuvcross}
\end{figure*}

\begin{figure*}
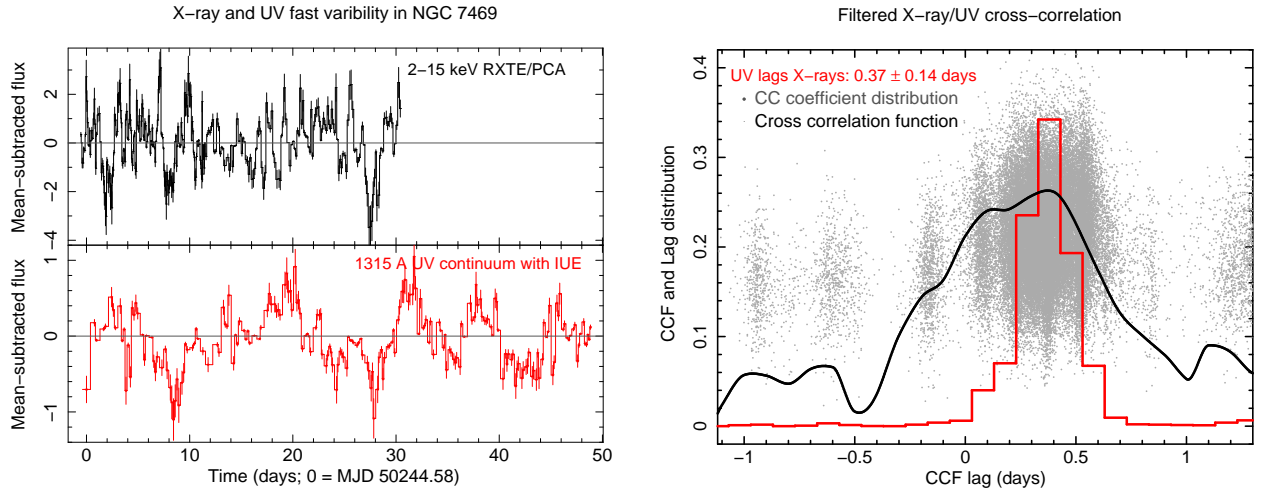

\centering
\includegraphics[scale=0.33,angle=-90]{fig3a.ps}
\includegraphics[scale=0.33,angle=-90]{fig3b.ps}
\caption{{\it The effect of variability filtering:} After filtering the variability longer than 5 days from X-ray and UV lightcurves shown in the left panel of Figure \ref{xuvcross}, the resulting lightcurves are shown in the left panel. When these filtered lightcurves are cross-correlated using the MC based flux randomisation and random subset selection methods combined (FR+RSS), the resulting frequency distribution is shown in the right panel where a switch in lag is observed at a timescale comparable to the accretion disc reverberation timescale. The solid line shows the interpolated cross correlation function (CCF). The grey dots are the centroid lags from all of the FR+RSS MC based centroid lag calculations and the histogram give the $\tau_{cent}$ distribution for dots above the adopted threshold, CCF$>$0.2.}
\label{xuvcross2}
\end{figure*}

\begin{figure*}
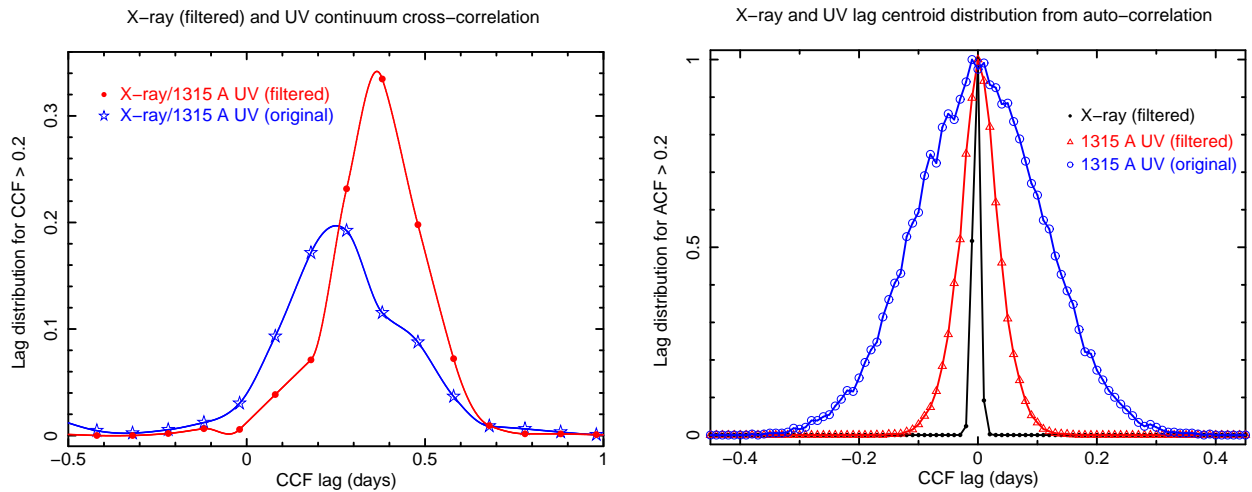

\centering
\includegraphics[scale=0.33,angle=-90]{fig4a.ps}
\includegraphics[scale=0.33,angle=-90]{fig4b.ps}
\caption{{\it Effect of filtering on X-ray/UV cross correlation:} Left panel shows the lag distribution for FR/RSS samples with centroid CCF $>$ 0.2 when the hi-pass filtered X-ray lightcurve is cross-correlated with the filtered UV continuum at 1315~\AA ~(circles) and with the unfiltered UV continuum (stars) using the flux randomisation (FR) and random subset selection (RSS) methods combined. A significant overlap between the two distributions can be observed. Right panel shows auto-correlation distribution function of the filtered X-ray (filled circles) and both filtered (triangles) and unfiltered (empty circles) UV continuum at 1315~\AA. Both filtered and unfiltered UV auto-correlation functions are wider than filtered X-ray auto-correlation function.}
\label{xuvauto}
\end{figure*}

\section{Observation}

\xte{} performed 311 observations of NGC~7469 between 10 June 1996 00:44:16 and 11 July 1996 23:59:19. For each observation, we extract the \xte{}/PCA lightcurve in the 2-15 keV energy range, combining observations from PCU0, PCU1 and PCU2 which were operational during the entire period of the observations. 
\iue{} continuously monitored NGC~7469 between 1996 June 10 and July 29 producing in total of 218 low-dispersion UV spectra in the wavelength range of 1150-1975~\AA. The details of the analysis procedure are provided in \citet{wa97}. Raw images were processed using the TOMSIPS \citep{ay93} and NEWSIPS \citep{ni93} data reduction packages. In this work, we consider NEWSIPS pipeline reduced spectra as mentioned by \citet{wa97}, a nonlinear wavelength calibration error exists in TOMSIPS reduction since long-term drifts in the wavelength scale were not taken into account. On the other hand, NEWSIPS reduced spectra matches well with the \hst{} spectra without any corrections applied. 
We also use data from \swift{} and the Wise Observatory optical telescope which are described later. Details of observations which are used to perform continuum reverberation mapping are provided in Table \ref{obs}.

\section{Data Reduction}
Lightcurve extractions are performed using {\tt Heasoft 6.25} packages applying standard filtering criteria. Further details on the observation and analysis procedures are provided by \citet{na98}.

In case of NEWSIPS pipeline reduced \iue{} spectra, due to a small shift (1-2~\AA) in wavelength caused by the large aperture pointing errors \citep{wa97}, an offset compensation is performed so that the sharp \civ{} line feature of all spectra falls at same average wavelength. Extinction corrections are not significant due to the very low interstellar reddening E(B-V) = 0.059 and background corrections are applied. The red-shift observed from the \civ{} average peak is consistent with the spectroscopic red-shift of $z=0.00163$ from the [OIII] line at 5007 \AA~ \citep{sa95}.  

\subsection{UV spectral fitting}
\iue{} UV spectra show significant variability in the continuum flux level. An example of such variations is shown in the left panel of Figure \ref{iuespec}. At all wavelengths between 1220-1970 \AA, the continuum UV flux as observed on 1996 June 15 is significantly higher than that on 1996 June 19. To study flux variability, we extract lightcurves at  different wavelengths.  
To extract UV continuum and line lightcurves from UV spectra, we adopt a slightly different approach than \citet{wa97}. Using the $\chi^2$ minimisation technique in {\tt XSpec}, we fit the average spectrum with suitable combinations of a powerlaw function that represents the underlying continuum and multiple narrow and broad Gaussian components that describe emission lines. The best-fit model yields residuals with $\chi^2$ per degrees of freedom = 179/175. The right panel of Figure \ref{iuespec} shows the time-averaged UV spectrum fitted with different model components and the residual of the fitting. The powerlaw energy spectral index from the best-fit average UV spectrum is observed to be -1.67 $\pm$ 0.11 which is consistent with that typically observed from AGN \citep{sh12} and also consistent with the radio-quiet nature of NGC 7469 \citep{ba15}. Using the best-fit model, we fit individual \iue{} spectra by fixing the continuum powerlaw index and letting all other parameters vary. From individual spectrum fitting, we derive spectral line and continuum parameters such as line width and flux. One advantage of using spectral modelling over numerical integration is that line and continuum fluxes can be measured more accurately, particularly when narrow/broad lines are close to each other and there exists an underlying broad continuum. For example, \citet{ve01} showed that the asymmetry in the blue wing of the \civ{} line is caused by the presence of high-ionization {Si~{\sc ii}} lines at $\sim$1540~\AA. Therefore, the use of two Gaussian components at both line locations provide more accurate modelling and hence flux measurements of the underlying continuum. From the best-fit model, continuum UV flux is computed in the rest-frame wavelength range of 1306-1327~\AA, 1473-1495~\AA, 1730-1750~\AA ~and 1805-1835~\AA ~using a convolution model ({\tt cflux} in {\tt XSpec}) that provides integrated, continuum-subtracted flux over a given wavelength range and its 1$\sigma$ error. Wavelength ranges for continuum flux measurement are kept consistent with those from \citet{wa97}.

\subsection{Filtering}

In this paper we are searching for signs of reprocessing in the original \xte{} and \iue{} observations, which are dominated by a very small number (2 in the X-ray observations, 3 in the \iue{} campaign) of large amplitude variations with peak to trough timescales of $\sim10$ days. If NGC~7469 behaves like other AGN of similar mass and accretion rate, the reprocessing signature should be manifest by the UV lagging the X-ray lightcurves by less than a day. It is well established that long-term variations can distort the measurement of short-term lags in CCFs \citep{we99} and so we filter out the long timescale, large amplitude, variations. We choose a 5-day filtering timescale which will eliminate variations on longer timescales but will allow lags on timescales shorter than 5 days, of both positive (i.e., reprocessing) and negative (seed photon variation) sign, to be detected. Similar filtering techniques have been used successfully to reveal short timescale correlations in other AGN, eg NGC~5548 \citep{mc14} and NGC~4593 \citep{mc18}.

To filter the lightcurves, we use a locally-weighted scatter plot smoothing (LOWESS) function which is based upon a non-parametric, non-linear least square regression method \citep{cl88}. The weight function used for LOWESS is the tricube kernel function: $k(d) = (1-|d|^3)^3$ where $d$ is the distance of a given data point from the point on the curve being fitted, scaled to lie in the range from 0 to 1. For the filtering purpose, $d(t)\equiv(t-t_i)/\Delta t$ is the time difference between time $t$ and the data point $i$ at time $t_i$ in units of $\Delta t=5$~d. Such a function has higher efficiency than traditional kernel functions like boxcar or triangular and does not require specification of the model function to fit the data, therefore, making it ideal to fit complex processes where no theoretical model exists. When compared to the efficiency of the Epanechnikov kernel function, the relative efficiency of the LOWESS kernel is 99.8\% while the same for the boxcar and triangular kernels are 92.9\% and 98.6\% respectively \citep{ep69}. The efficiency of a function f(x) is defined as $\sqrt{\int x^2 f(x) dx}$  $\int{f(x)^2 \mathrm{d}x}$. 

In this work, the LOWESS filter is used (residual from the LOWESS function fitting) to eliminate variability slower than 5 days in the X-ray and UV lightcurves. In the rest of the paper and in all Figures, the word `filtered' implies a 5-day filtering unless otherwise specified.  

\subsection{Correlation and delay measurements}

Since X-ray and UV observations have dissimilar temporal coverage, we use the interpolated cross-correlation function (CCF) to compute the delay among X-ray and various UV continuum and line lightcurves. Uncertainties on lag measurements are computed using a Monte Carlo simulation to assess the flux uncertainties associated with each measurement and the sampling uncertainties of the observed time series, similar to bootstrapping. The details of the implementation of both methods are provided in \citet{pe98} and denoted as the flux randomisation (FR) and random subset selection (RSS) methods respectively. For each Monte Carlo realisation of a light curve with N data points, N selections are drawn at random (i.e., random subset selection or RSS); for data points that are randomly selected multiple times $M$, the associated uncertainty is decreased by $M^{1/2}$. The data points are then altered by adding random Gaussian deviates with a dispersion equal to the assigned flux uncertainty (i.e., FR). Multiple realisations result in a distribution of cross-correlation functions, and the corresponding joint distribution of CCF peak correlation coefficients and centroid lags. Centroid lag ($\tau_{cent}$) is computed by averaging lag over regions where CCF is above 80\% of the peak.

We perform 100,000 FR+RSS simulations for each pair of X-ray/UV lightcurves in this work and consider only measurements for which the cross-correlation coefficient is higher than 0.2. Although the resultant lag distribution may deviate from the normal distribution, we fit the resulting CCF centroid distribution using a Gaussian function and interpret the Gaussian centre as the measured time delay and its half width at half maxima (FWHM/2) as the uncertainty on the delay measurements. 
\begin{figure*}
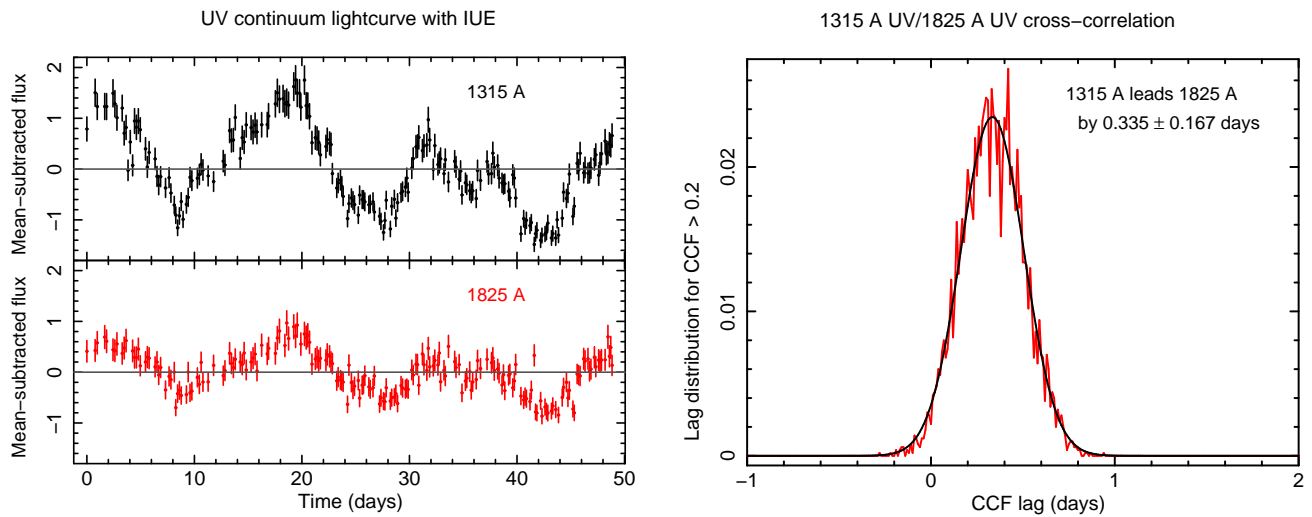

\centering
\includegraphics[scale=0.34,angle=-90]{fig5a.ps}
\includegraphics[scale=0.34,angle=-90]{fig5b.ps}
\caption{{\it UV continuum lightcurve and CCF:} Left panel shows 1315~\AA~(top) and 1825~\AA~(bottom) UV continuum lightcurves from NGC 7469. Both lightcurves are mean-subtracted residuals and similar y-axis scales are used for easy comparison of their flux variability. When they are cross-correlated using FR+RSS method, the distribution of centroid lags (for CCF $>$ 0.2) is shown in the right panel. The frequency distribution is fitted with a Gaussian whose centroid and the FWHM/2 are quoted as the time delay and its uncertainty respectively.}
\label{uvcontlight}
\end{figure*}

\begin{table}
    \centering
    \begin{tabular}{c|c|c|c}
    \hline
         Telescope/ & Observation & no. of exposures & Wavelength\\
         Instrument & time (MJD) & /data points & coverage (\AA)\\
    \hline
         RXTE/PCA & 50244-50275 & 311 & 0.83-6.22 \\ 
         IUE & 50244-50294 & 218 & 1150-1975 \\
         Swift/XRT & 54630-58230 & 176 & 1.24-24.8 \\ \hdashline
         Swift/UVOT &  &  & \\ 
         {\it UVW2} & 54644-58229 & 98 & 1928$\pm$657 \\
         {\it U} & 54635-57423 & 61 & 3465$\pm$785 \\
         {\it UV-GRISM} & 56410-56524 & 37 & 1700-2900 \\ \hdashline
         $^c$Wise/FOSC & 50237-50295 & 42 & 4016-7841 \\ 
    \hline
    \multicolumn{4}{l}{$^c$ AGN Watch campaign}
    \end{tabular}
    \caption{Details of observations from different telescopes and satellites used in the present work. Central wavelengths and FWHM \citep{po08} are quoted in case of UVW2 and U filters.}
    \label{obs}
\end{table}

\section{Analysis and results}
The left panel of Figure \ref{xuvcross} shows the 2-15 keV X-ray lightcurve (top) and the 1306-1327~\AA~UV continuum lightcurve (bottom; referred here 1315~\AA~band) observed with \xte{} and \iue{} respectively. Both lightcurves are shown as residuals after subtracting the mean flux for comparison of lightcurve variability relative to their mean value. The right panel shows the centroid lag distribution of the FR+RSS cross-correlation between the X-ray and UV continuum lightcurves. The cross-correlation distribution peak indicates that X-rays are delayed relative to the UV 1315~\AA~band continuum by 3.49 $\pm$ 0.22 days. This is consistent with previous measurements by \citet{na98} but inconsistent with reprocessing models in which the UV variations should lag the X-rays.

\subsection{Effect of lightcurve filtering}
The filtered X-ray and UV lightcurves are shown in the left panel of Figure \ref{xuvcross2}. Visually the lightcurves are now quite similar. 
An FR+RSS cross-correlation between filtered lightcurves along with the cross correlation function, shown in the right panel of Figure \ref{xuvcross2}, clearly demonstrates that on a timescale faster than 5 days, the UV continuum lags the X-rays by 0.37$\pm$0.14 days. 
The solid black line in the right hand panel of Figure \ref{xuvcross2} shows the interpolated cross correlation function \citep{wh94} for the lightcurves in the left hand panel of Figure \ref{xuvcross2}. The grey dots are the centroid lags from all of the FR+RSS MC based centroid lag calculations based on these same lightcurves. The histogram is the distribution of MC centroid lags selecting only those where peak CCF value is $>$ 0.2.
A similar lag is obtained when we truncate the UV lightcurve to match the duration of the X-ray lightcurve. Such a lag timescale is consistent with the accretion disc reverberation delay observed from other AGN (see sect. 5 and 6 for details). Such a switch in the sign of the lag, as well as the change in lag timescale, is remarkable. 

To check the effect of UV filtering on the measured cross-correlation with X-rays, we perform the FR+RSS cross-correlation between the filtered X-ray and unfiltered UV lightcurve and plot the CCF lag distribution in the left panel of Figure \ref{xuvauto} along with the X-ray/UV CCF centroid distribution with the filtered UV continuum. Clearly, X-ray and UV CCF centroid distributions using both filtered and unfiltered UV continuum significantly overlap with each other. Such an overlap implies that both fast and slow UV variability from hours to days timescale is mostly driven by fast X-ray variability.

To explore the connection between the driving and driven variability further, we perform auto-correlation analysis using filtered X-ray, filtered and unfiltered UV lightcurves. The FR+RSS method and the auto-correlation distribution is shown in the right panel of Figure \ref{xuvauto}. The lag centroid distribution of all three ACFs are well defined. However, the X-ray auto-correlation is significantly narrower than both filtered and unfiltered 1315~\AA~UV auto-correlation. Such characteristics indicate that the UV variations are smoother than that of the X-ray variability, causing a wider ACF lag distribution for UV than for X-rays.
\begin{figure*}
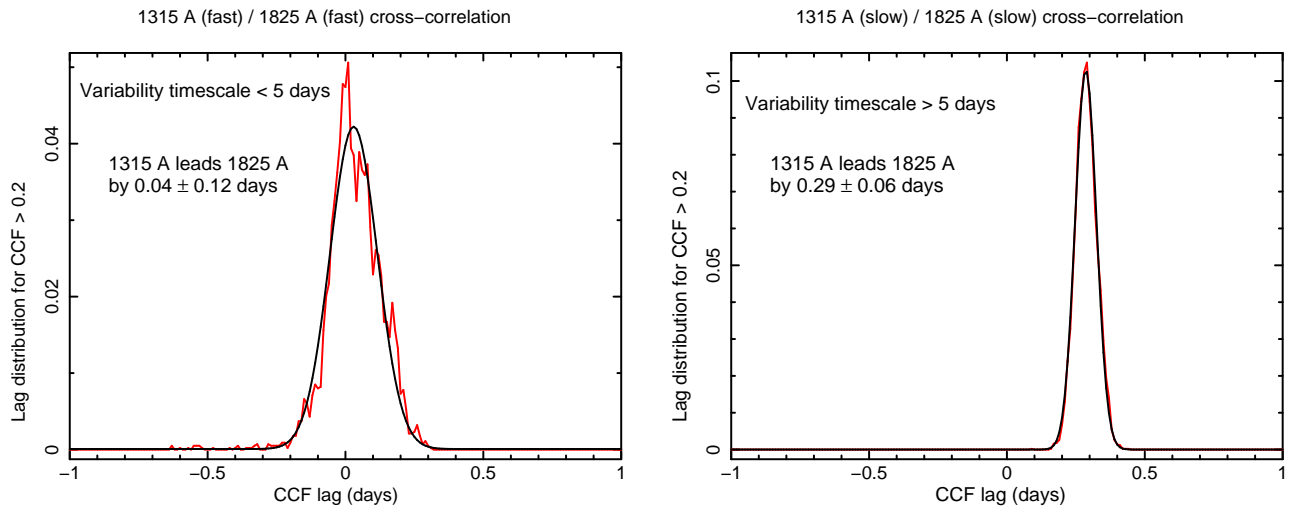

\centering
\includegraphics[scale=0.34,angle=-90]{fig6a.ps}
\includegraphics[scale=0.34,angle=-90]{fig6b.ps}
\caption{{\it The effect of lightcurve filtering in UV:} Left panel shows relative frequency distribution of CCFs computed with 1315~\AA~and 1825~\AA~UV continuum lightcurves (left panel of Figure \ref{uvcontlight}) after applying a slow filter which only preserve fast (faster than 5 days) variability. On the other hand, when a fast filter is applied which only preserve slow (slower than 5 days) variability, the resulting CCF is shown in the right panel. Both distributions is fitted with a Gaussian whose centroid and FWHM/2 are quoted as the time delay and its uncertainty respectively. A significant difference between the fast and slow variability delay measurements can be noted. }
\label{uvcontfilt}
\end{figure*}

To test whether the delays among different UV bands depend on wavelength, we carry out a further check. We consider 1315~\AA ~and 1825~\AA ~(integrated flux in 1805-1835~\AA) band continuum lightcurves shown in the left panel of Figure \ref{uvcontlight}. Both lightcurves have similar coverage, the same number of data points and the variability at different timescales is most likely driven by similar physical processes. Both lightcurves are mean-subtracted residuals and similar y-axis scales are used for an easy visual comparison of their variability. A cross-correlation between both continuum bands indicates that 1825~\AA ~continuum lags behind the 1315~\AA ~continuum by 0.34 $\pm$ 0.17 days, as shown in the right panel of Figure \ref{uvcontlight}. We apply the filter to both lightcurves, decomposing them into slow and fast components separated at a 5-day timescale. The cross-correlation between the 1315~\AA ~and 1825~\AA ~fast lightcurves using FR+RSS technique yields that the 1825~\AA ~fast variability lags behind the 1315~\AA ~fast variability by 0.04 $\pm$ 0.12 days, consistent with zero lag and shown in the left panel of Figure \ref{uvcontfilt}. However, when the 1315~\AA ~slow variability is cross-correlated with the 1825~\AA ~slow variability, the resulting distribution indicates that 1825~\AA ~band is delayed to the 1315~\AA ~band by 0.29 $\pm$ 0.06 days (shown in the right panel of Figure \ref{uvcontfilt}). Therefore, the slower variability is significantly more delayed than the faster variability in the longer UV continuum.

\section{Swift monitoring and UV/optical lag measurements}
\begin{figure*}
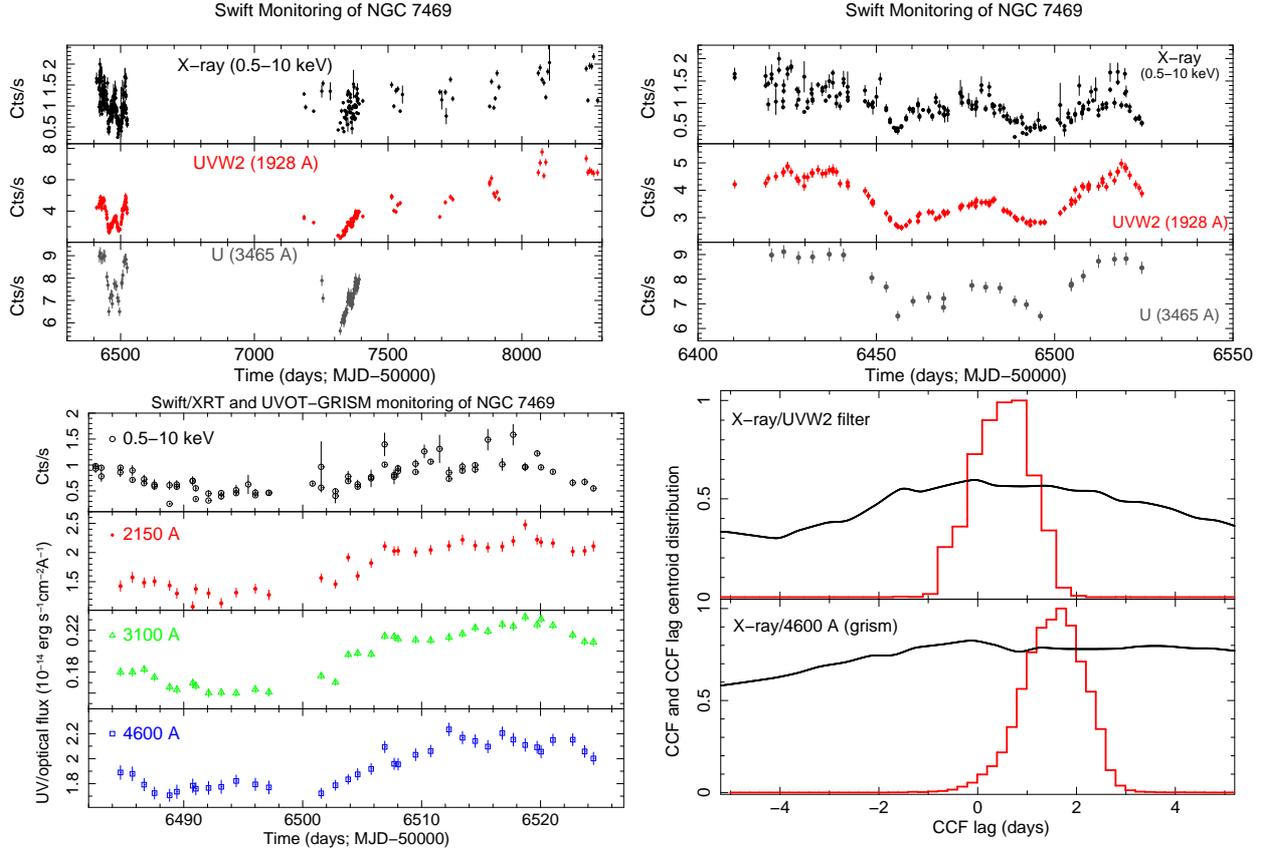

\centering
\includegraphics[scale=0.33,angle=-90]{fig7a.ps}
\includegraphics[scale=0.33,angle=-90]{fig7b.ps}
\includegraphics[scale=0.33,angle=-90]{fig7c.ps}
\includegraphics[scale=0.32,angle=-90]{fig7d.ps}
\caption{{\it Reverberation delay measurements from X-ray to optical} Top left panel shows long-term ($\sim$6 years) lightcurve of NGC 7649 as monitored by \swift{}/XRT 0.5-10 keV X-ray (top) and \swift{}/UVOT UVW2 filter at 1928 \AA (middle) and U filter at 3465 \AA (bottom). To have a visual clarity of the correlated variability among X-ray, UVW2 and U filters, a zoomed version of the same lightcurve between MJD 56400 and MJD 56550 is shown in the top right panel. During the same \swift{} campaign, continuum lightcurves obtained from \swift{}/GRISM observations at 2150~\AA, 3100~\AA~and 4600~\AA~are shown in the bottom left panel along with X-ray lightcurve. Bottom right panel shows cross correlation function and CCF centroid distribution from FR+RSS between X-ray and UVW2 filter lightcurves (top) and X-ray and 4600~\AA~GRISM lightcurve (bottom) respectively.}
\label{swift}
\end{figure*}

\begin{figure*}
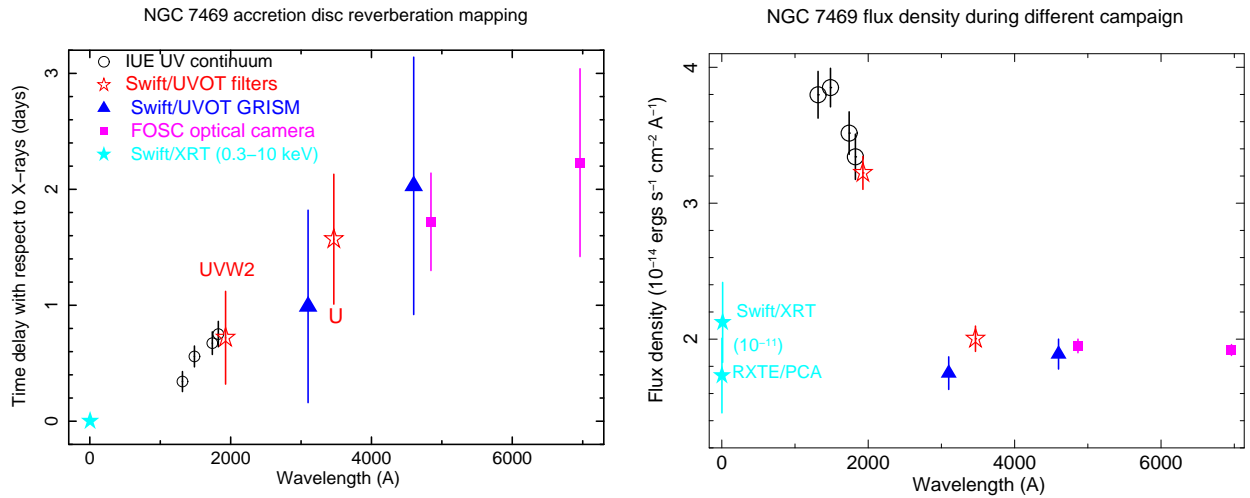

\centering
\includegraphics[scale=0.33,angle=-90]{fig8a.ps}
\includegraphics[scale=0.33,angle=-90]{fig8b.ps}
\caption{{\it Reverberation delay measurements from X-ray to optical:} Wavelength-dependent time delay of different UV filter and GRISM lightcurves with respect to the simultaneous X-ray lightcurve are shown by stars and triangles respectively in the left panel. For the sake of completeness, we also include delay measurements at longer wavelength performed by the FOSC spectroscopic camera on 1m optical telescope at the Wise Observatory partly simultaneous with the \xte{} observations. Optical measurements are shown by squares. Empty circle are lag measurements using \iue{} continuum. Flux density measurements during different multi-wavelength campaigns are shown in the right panel. X-ray flux densities are shown in the unit of 10$^{-11}$ for easy comparison. Different symbols in this panel have the same meaning as in the left panel. }
\label{swift2}
\end{figure*}

The log of \swift{} observations is given in Table~\ref{obs}.
There are 176 XRT photon counting X-ray (0.5-10 keV) visits. Most of the X-ray observations were accompanied by UVOT imaging mode UVW2 (1928~\AA) observations with a lesser number including U (3465~\AA) observations. Some UVOT observations employed the GRISM.
X-ray and UV flux measurements were performed using the Southampton pipeline\footnote{\url{https://swiftly.soton.ac.uk/}} \citep{mc18} based on standard \swift{} data analysis procedures \citep{ca12}. The X-ray, UVW2, U filter lightcurves of NGC~7469 are shown in the top left panel of Figure \ref{swift}. On two occasions, \swift{} observed NGC~7469 continuously over a period of nearly 4 months in X-ray, UVW2 and U filters. One such set of observations taken between MJD 56400 and MJD 56550 is shown in the top right panel of Figure \ref{swift}. We have also extracted continuum lightcurves from 37 observations taken using \swift{}/GRISM between 28 April and 20 August 2013. Using GRISM data, continuum fluxes are measured at 2150~\AA, 3100~\AA~ and 4600~\AA~ respectively, and corresponding lightcurves are shown in the bottom left panel of Figure \ref{swift}. For each UVOT filter and each continuum lightcurve from GRISM, we perform FR+RSS cross correlation with the X-ray lightcurve to measure the wavelength dependent lag. The bottom right panel of Figure \ref{swift} shows the CCF and corresponding lag distribution computed for X-ray and UVW2 filter (top) and for the X-ray and GRISM (bottom), respectively. 
We performed a wavelength-dependent cross correlation study between the X-ray and UVW2, U filter lightcurves using the FR/RSS technique. With respect to the X-rays, we found that the UVW2 and U filter lightcurves are delayed by 0.72 $\pm$ 0.51 days and 1.57 $\pm$ 0.71 days, respectively. These delays are shown by empty stars in the left panel of Figure \ref{swift2} while the X-ray point is shown by the solid star. Along with the \swift{} measurements, for comparison, we include the continuum UV delay from \iue{} (shown by empty circles). The lag uncertainty merely reflect the number of data points and the measurement errors in the relevant lightcurves.

 We performed the FR+RSS cross-correlation between 2150~\AA~and~3100~\AA~and between~2150~\AA~and 4600~\AA~lightcurves respectively and measured the delay. To verify and confirm our results, we repeat the similar exercise of the delay measurement by replacing the 2150~\AA~lightcurve with the simultaneous \swift{}/UVW2 lightcurve. They are similar to within the measurement uncertainties. The Resulting delays with respect to X-rays are shown by the triangles in Figure \ref{swift2}. To compare fluxes during different campaigns, we compute and plot the average flux density at different wavelengths in the right panel of the Figure \ref{swift2}. Flux densities at similar wavelengths are consistent during different campaigns.
 
\subsection{Optical continuum from FOSC spectroscopic camera}
During the \xte{} and \iue{} joint campaign in 1996, NGC~7469 was also monitored using the Faint Object Spectroscopic Camera (FOSC) mounted on the 1m optical telescope at the Wise Observatory, Tel Aviv University \citep{ka96,co98}. Between 02 June and 30 July 1996, 42 spectroscopic observations were taken with a spectral resolution of $\sim$6 \AA in the wavelength range 4016-7841 \AA \citep{co98}. While the analysis details and lightcurves are provided by \citet{co98}, the optical continuum lightcurves at 4845 \AA~ and 6962~\AA~ obtained from the campaign\footnote{\url{http://www.astronomy.ohio-state.edu/~agnwatch/n7469/lcv/}} are cross-correlated with respect to the 1300~\AA~ UV continuum lightcurve from \iue{} using FR+RSS technique and the resulting delays are shown by squares in Figure \ref{swift}.

\section{Wavelength-dependent lag modelling}
To understand the nature of the observed wavelength-dependent reverberation delay and test the compatibility with the prediction of standard accretion disc theory as outlined in Section 1, i.e., $\tau \propto \lambda^{4/3}$, we performed modelling using two approaches. In the first approach, shown in the top panel of Figure \ref{model} we fit the wavelength-dependent delay using a power-law model first optimising both the normalisation and index (shown by the dotted line) and then with the index fixed at 4/3 (shown by the solid line). Whilst an index of 4/3 is an acceptable fit, the best-fit index is 0.89 $\pm$ 0.09. A similar index was noted by \citet{st17} in NGC~5548.  
Many earlier works \citep{ed17,mc18,ca18} showed that the X-ray delay is usually offset with respect to the standard disc theory prediction. Therefore, in the second approach, we fit the observed delay with an offset powerlaw (constant+powerlaw) where the offset and powerlaw normalisation are free to vary while the index is fixed to 4/3. The resulting fit is shown in the bottom panel of Figure \ref{model}. According to the best-fit model the X-ray delay is offset by $\sim$0.38 day. Interestingly, the 1300 \AA\ UV continuum delay is also offset by $\sim$0.1 day from the best-fit prediction. Assuming the lamp-post geometry of the corona, the reverberation delays at different wavelengths are calculated with respect to X-rays and shown by blue triangles with the dotted line. Details of the calculation are provided in the next section.  

\section{Discussion and Conclusions}

In this work, we fit individual \iue{} 1220-1970 \AA~ spectra and extract the continuum UV lightcurves from the model fitted parameters. Using the Monte Carlo simulation-based cross-correlation techniques, we show that the 2-10 keV X-ray lightcurve from \xte{} lags behind the UV continuum lightcurve by 3.49$\pm$0.22~days (Figure \ref{xuvcross2}). However, if we filter out variability slower than 5 days from the X-ray lightcurve, the cross-correlation shows that UV variability lags the X-ray variability by 0.37$\pm$0.14 days. The UV lag is consistent with the same value for both filtered and unfiltered UV lightcurves. Such a delay timescale is consistent with the light travel time from the X-ray emitting corona to the UV emitting region in the accretion disc and therefore fully consistent with the accretion disc reprocessing scenario. Therefore, UV continuum variability, from hours to weeks timescale is mainly driven by the short-term, large X-ray variability. 
\subsection{Evidence for finite size and temperature gradient in the reprocessing region}
Between the 1315~\AA~and 1825~\AA~ UV continuum lightcurves, we show (Figure \ref{uvcontfilt}) that slower ($>$5d) variability is delayed ($0.29 \pm 0.06$~day) while the faster ($<$5d) variability is not ($0.04 \pm 0.12$~day). This is consistent with the accretion disc origin of UV variability at a different wavelengths.
Also, the auto-correlation function is broader for UV than for X-rays (Figure \ref{xuvauto}). Both results, delayed slow variations and broad ACF, point to UV reprocessing from an extended rather than compact region. The outer region of the reprocessed area produces slower variability while for the same wavelength emission, the inner region causes faster variability. The idea that reprocessing occurs from a hotter or larger disc was provided by \citet{mc14,fa16,ha18,ka19} and also supported by the microlensing observations \citep{mo10}. If the delayed, reprocessed, UV continuum at a particular wavelength originates from a narrow region of the accretion disc, we would expect the lag distribution to be narrow and symmetric, and the UV ACF width to be similar to that of the X-rays. 

\subsection{Role of fast X-ray variability}
\begin{figure*}
\centering
\includegraphics[scale=0.46]{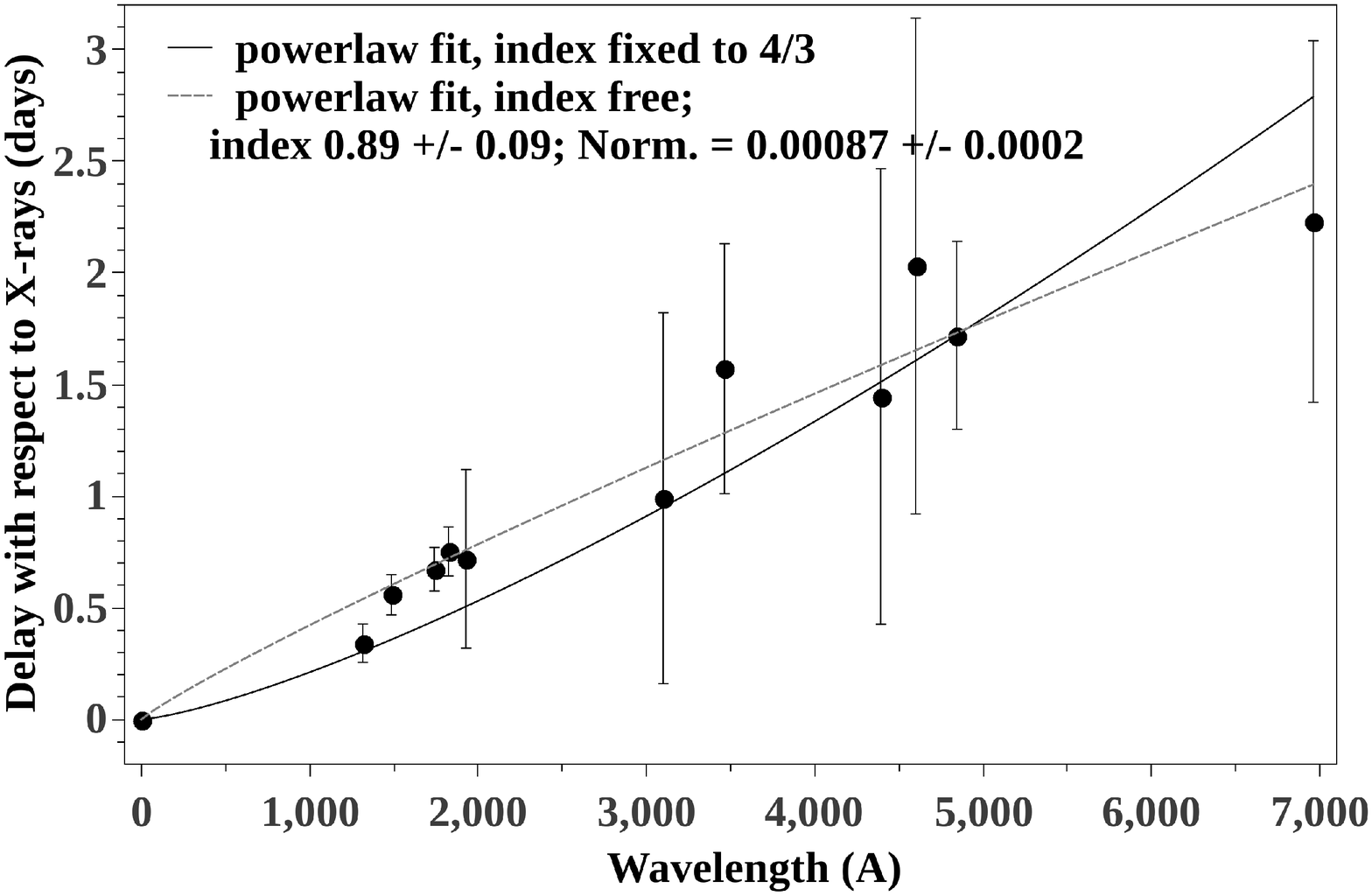}
\includegraphics[scale=0.43]{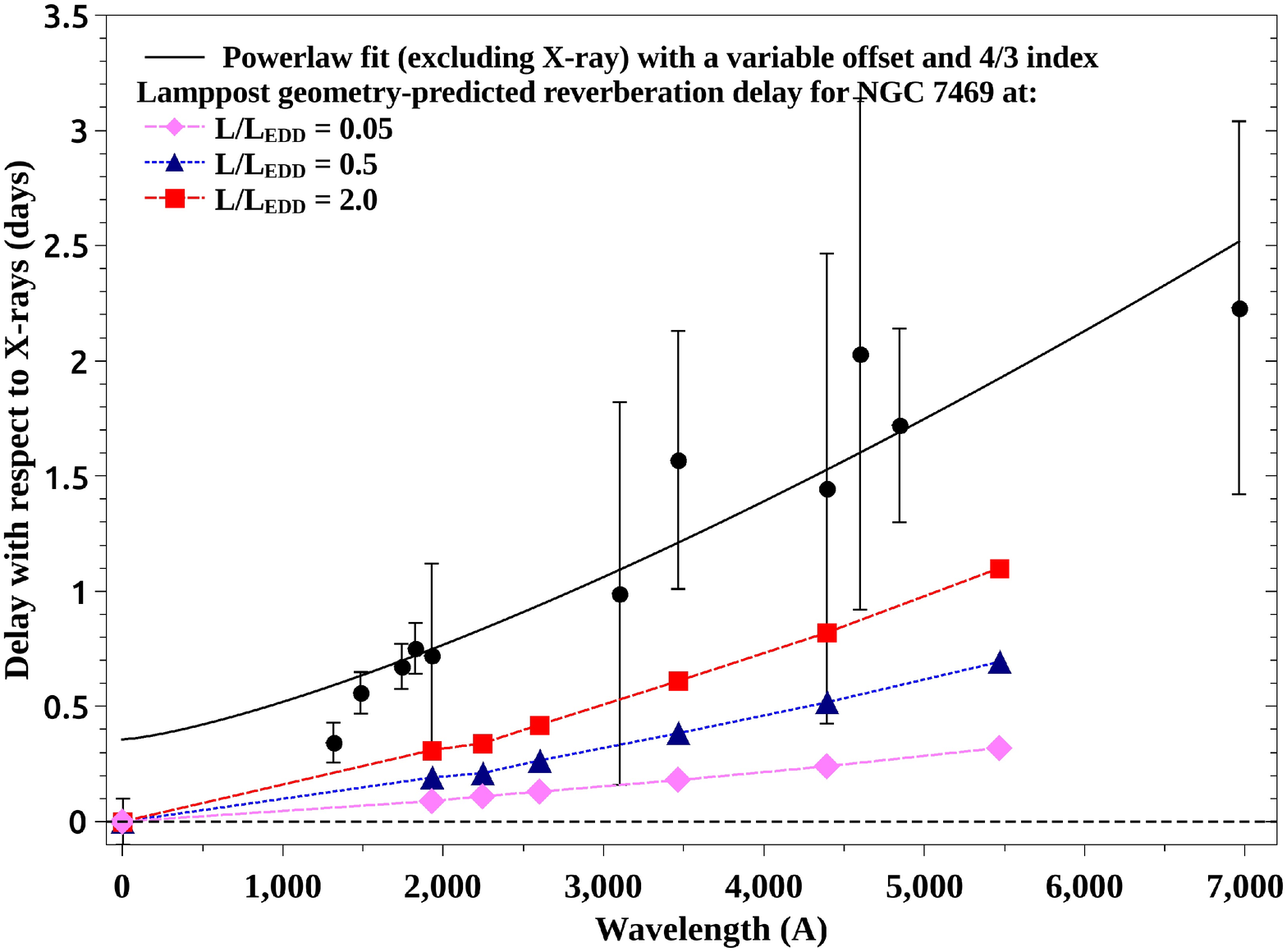}
\caption{{\it Reverberation delay modelling from X-ray to optical:} Top panel: the fitting of wavelength-dependent delay spectrum (same spectral points as the left panel of Figure \ref{swift2} but all represented by solid circles for the sake of uniformity) including X-rays using two powerlaws: with variable normalisation and index (shown by the dotted line) and with variable normalisation but fixing the index at 4/3 (shown by the solid line). Bottom panel: the powerlaw fit of wavelength-dependent delay with an offset where the offset and the powerlaw normalisation are free to vary while the index is fixed to 4/3. The X-ray data point is excluded while fitting and shown for clarity. The best-fit shows the X-ray and UV (1300~\AA) delay measurements are below the fitted model by $\sim$0.38 day and $\sim$0.1 day respectively. Squares, triangles and diamonds with the dotted lines show the theoretical X-ray reverberation delay estimation with $L/L_{EDD}$ of 0.05, 0.5 and 2.0 respectively for NGC 7469 assuming the `lamppost' geometric configuration of the corona without any additional X-ray offset.}
\label{model}
\end{figure*}

Wavelength-dependent delay analysis using filtered X-ray and UV continuum lightcurves is found to be consistent with the predicted delay due to the reprocessed UV emission from the geometrically thin and optically thick accretion disc (Figure \ref{model}). The inclusion of \swift{} delay measurements from NUV to optical is also consistent with the standard disc reprocessing delay, $\tau \propto \lambda^{4/3}$ relationship. Therefore, our study supports the hypothesis that the reverberation delay in the AGN accretion disc from NUV to optical is mostly driven by the X-ray variability faster than a week. The X-ray variability in AGN is usually associated with the size of the corona, which in turns depends upon the central black hole mass and in a sample of Seyfert 1 galaxies. \citet{lu01} showed that the excess variance of the short timescale X-ray variability is anti-correlated with the black hole mass. Therefore, it suggests that in all AGN that show reverberation continuum delays, the UV delay is driven by the fast X-ray variability, but the fast X-ray variability timescale may vary depending upon the central black hole mass. However, testing such a hypothesis is beyond the scope of the present work.

\subsection{Comparison with the theoretical prediction}
We have compared the measured lags (Figure \ref{swift2}) with those expected following illumination of just an accretion disc by a point X-ray source located $6\,R_{\rm g}$ above the spin axis of the black hole. We use the same model as \citet{mc18}, i.e., we derive the temperature distribution around a black hole of a smooth accretion disc of the form described by \citet{sh73}. We then illuminate the disc with X-ray impulse illumination and, taking into account the resultant change in surface temperature distribution, calculate the response in various UV and optical wavebands. The X-ray impulse response is computed at six \swift{}/UVOT filter wavelengths due to the availability of their filter response curves. 

As in previous work, we take the lag as the time for half of the reprocessed light to arrive (see \citet{mc18} for a discussion of this point).
We consider a Schwarzchild black hole and an inclination of the disc of 45 degrees. We adopt a black hole mass 
$9 \times 10^{6}$ \(\textup{M}_\odot\) \citep{pe14a} for which $L_{\rm Edd}$ = $1.13 \times 10^{45}$ erg~s$^{-1}$.
We take the illuminating X-ray luminosity from the \swift{}/BAT 70 month survey \citep{ba13} of 1.8 $\times$ 10$^{43}$ ergs/s and multiplied it by a factor of 2 to extrapolate from the observed 14-195 keV to a broader 0.1-500 keV band. The exact value of this parameter is not critical. The accretion rate is not well known as a central starburst ring contaminates the bolometric luminosity. Values of \me{} ~between 0.05 \citep{mh18} and 2 \citep{wo02} have been quoted. We note that the total X-ray luminosity is then 3\% of the Eddington luminosity. Assuming even a very modest X-ray to bolometric correction of a factor 10 \citep{ne19}, we derive an accretion rate of \me$\sim0.3$ and most correction factors are larger than that value. We there take \me{}~=0.5. 
\subsubsection{Simulation results and inference}
In the top panel of Figure~\ref{model} we plot the model theoretical values assuming \me{}~=0.5 but also show model lines covering two extremes \me{} of 0.05 and 2. Here we plot lags relative to the X-ray band. However although the lags relative to the \swift{} UVW2 band generally follow a smooth curve, and are similar in most AGN, the lag of the UVW2 relative to the X-rays is usually much larger than expected from an extrapolation of the longer wavelength lags down to the X-ray band \citep{mc18}. Here we note similar effects. The lag spectrum dips down below a simple powerlaw fit at wavelengths shorter than 2000\AA and the model lag between the UVW2 and V-band (0.69 day) is a factor 2.3 less than the observed lag between UVW2 and V-band derived from the simple 4/3 powerlaw fit to the data in Figure~\ref{model}. 

Discrepancy between the observed and model lag by a similar factor just within the optical bands was first noted by \citet{co99} and within the UV and optical bands by \citet{mc14}. As we have no observed value of the V-band lag, this simple model fit is our best estimate of an observed lag. A factor of 2.3 is close to the average ratio of model to observed UVW2 to V-band lags in other AGN \citep{mc18}.
If we instead chose a value of \me{}~=2, the ratio between observed and model lag would drop to 1.45 and if we chose a value of \me{}~=0.05, the ratio would rise to 5. Although the lags to the longer wavelength bands are not measured here to very high precision, nonetheless a factor of 5 discrepancy is significantly more than seen in other AGN whereas a factor of 1.45 would not be too different  - assuming they are also all Schwarzchild black holes with disc inclinations and illuminating source heights similar to those assumed here, which are significant assumptions. We conclude that NGC~7469 does have a high accretion rate, nearer to 50 percent than 5 percent.

We may note that the above calculation assumes the delay due to the reprocessed emission in the continuum band is due to the X-ray-heated accretion disc. However, several works \citep{ko01,ko19,la18,ch19} showed that the diffuse continuum (DC) emission from the extended BLR cloud may have significant contamination in the UV-optical continuum lag measurements, particularly close to the Balmer continuum. The detail study of the DC contribution to the observed lag in NGC 7469 is beyond the scope of the present work.

\section{Acknowledgement}  
We thank the referee for comments and suggestions which help to improve the clarity of the paper. MP acknowledges Royal Society-SERB Newton International Fellowship support funded jointly by the Royal Society, UK and the Science and Engineering Board of India (SERB) through Newton-Bhabha Fund. IMcH acknowledges support from a Royal Society Leverhulme Trust Research Fellowship LT160006 and from STFC grant ST/M001326/1. EMC gratefully acknowledges support from the National Science Foundation through award number AST-1909199. KH acknowledges support from STFC grant ST/R000824/1.

\bsp

\label{lastpage}

\end{document}